\PassOptionsToPackage{table,xcdraw,dvipsnames}{xcolor}
\documentclass[sigconf]{acmart}

\usepackage{booktabs}
\usepackage{arydshln}
\usepackage{multirow}
\usepackage{cancel}
\usepackage[normalem]{ulem}
\usepackage[htt]{hyphenat}
\usepackage{enumitem}
\usepackage{color, colortbl}

\newcommand{\charswap}{\texttt{NeighbCharSwap}}
\newcommand{\charsub}{\texttt{RandomCharSub}}
\newcommand{\charqwerty}{\texttt{QWERTYCharSub}}
\newcommand{\removestop}{\texttt{RemoveStopWords}}
\newcommand{\tfiverectotitle}{\texttt{T5DescToTitle}}
\newcommand{\orderswap}{\texttt{RandomOrderSwap}}
\newcommand{\backtranslation}{\texttt{BackTranslation}}
\newcommand{\tfiveqqp}{\texttt{T5QQP}}
\newcommand{\wordembedsyn}{\texttt{WordEmbedSynSwap}}
\newcommand{\wordnetsyn}{\texttt{WordNetSynSwap}}

\newcommand{\trecdl}{TREC-DL-2019}
\newcommand{\uqv}{UQV100}
\newcommand{\antique}{ANTIQUE}

\newcommand{\bm}{BM25}	
\newcommand{\bmrm}{RM3}
\newcommand{\KNRM}{KNRM}
\newcommand{\CKNRM}{CKNRM}
\newcommand{\EPIC}{EPIC}
\newcommand{\BERT}{BERT}
\newcommand{\T}{T5}

\newcommand{\genspec}{\emph{gen./specialization}}
\newcommand{\aspectchange}{\emph{aspect change}}
\newcommand{\misspelling}{\emph{misspelling}}	
\newcommand{\naturality}{\emph{naturality}}	
\newcommand{\paraphrase}{\emph{paraphrasing}}
\newcommand{\ordering}{\emph{ordering}}

\newcommand{\Genspec}{\emph{Gen./specialization}}
\newcommand{\Aspectchange}{\emph{Aspect change}}
\newcommand{\Misspelling}{\emph{Misspelling}}	
\newcommand{\Naturality}{\emph{Naturality}}	
\newcommand{\Paraphrase}{\emph{Paraphrasing}}
\newcommand{\Ordering}{\emph{Ordering}}

\newcommand{\rff}{RRF}
\newcommand{\oracle}{\emph{best query}}

\newcommand{\trad}{Trad}
\newcommand{\nn}{NN}
\newcommand{\nnlm}{TNN}

\newcommand{\one}{({\small I})}
\newcommand{\two}{({\small II})}
\newcommand{\three}{({\small III})}
\newcommand{\four}{({\small IV})}
\newcommand{\five}{({\small V})}
\newcommand{\six}{({\small VI})}

\newcommand{\querypair}{$\{q_i,q_j\}$}

\definecolor{myblue}{HTML}{544FF0}
\definecolor{mygreen}{HTML}{5CA31D}
\definecolor{myorange}{HTML}{F07946}

\AtBeginDocument{%
  \providecommand\BibTeX{{%
    \normalfont B\kern-0.5em{\scshape i\kern-0.25em b}\kern-0.8em\TeX}}}

\setcopyright{none}

\acmConference[Woodstock '18]{Woodstock '18: ACM Symposium on Neural
  Gaze Detection}{June 03--05, 2018}{Woodstock, NY}
\acmBooktitle{Woodstock '18: ACM Symposium on Neural Gaze Detection,
  June 03--05, 2018, Woodstock, NY}
\acmPrice{15.00}
\acmISBN{978-1-4503-XXXX-X/18/06}

\begin{document}

\title{Evaluating the Robustness of Retrieval Pipelines with Query Variation Generators}

\author{Gustavo Penha}
\email{g.penha-1@tudelft.nl}
\affiliation{%
  \institution{Delft University of Technology}  
  \city{Delft}  
  \country{Netherlands}  
}
\author{Arthur Câmara}
\email{a.barbosacamara@tudelft.nl}
\affiliation{%
  \institution{Delft University of Technology}  
  \city{Delft}  
  \country{Netherlands}  
}
\author{Claudia Hauff}
\email{c.hauff@tudelft.nl}
\affiliation{%
  \institution{Delft University of Technology}  
  \city{Delft}  
  \country{Netherlands}  
}


\begin{abstract}
Heavily pre-trained transformers for language modelling, such as BERT, have shown to be remarkably effective for Information Retrieval (IR) tasks. IR benchmarks evaluate the effectiveness of (neural) ranking models based on the premise that a single query is used to instantiate the underlying information need. However, previous research has shown that \one{} queries generated by users for a fixed information need are extremely variable and, in particular, \two{} neural models are brittle and often easily make mistakes when tested with adversarial examples, i.e. examples with minimal modifications that do not change its label. Motivated by those observations we aim to answer the following question with our work: \emph{how robust are retrieval pipelines with respect to different variations in queries that do not change the queries' semantics?} In order to obtain queries that are representative of users' querying variability, we first created a taxonomy based on the manual annotation of transformations occurring in a dataset (specifically \uqv{}) of user created query variations. For example, from the query `\textit{cures for a bald spot}' to the variation `\textit{cures for baldness}' we are applying a \emph{paraphrasing} transformation that replaces words with synonyms. For each syntax-changing category of our taxonomy, we employ different automatic methods that when applied to a query generate a query variation. We conduct experiments on two datasets (\trecdl{} and \antique{}) and create a total of 2430 query variations from 243 topics across both datasets. Our experimental results for two different IR tasks reveal that retrieval pipelines are not robust to query variations that maintain the content the same, with effectiveness drops of $\sim$20\% on average when compared with the original query as provided in the datasets. Our findings indicate that further work is required to make retrieval pipelines with neural ranking models more robust and that IR collections should include query variations, e.g. using the methods proposed here, for a single information need to better understand models capabilities. The code and datasets are available at \textcolor{RubineRed}{\url{https://github.com/Guzpenha/query\_variation\_generators}}.

\end{abstract}

\maketitle

\section{Introduction}

Heavily pre-trained transformers for language modeling such as BERT~\cite{devlin2019bert} have been shown to be remarkably effective for a wide range of Information Retrieval (IR) tasks~\cite{yang2019simple,nogueira2019passage,penha2020curriculum}. Commonly, IR benchmarks organized as part of TREC or other evaluation campaigns, evaluate the effectiveness of ranking models---neural or otherwise---based on small sets of topics and their corresponding relevance judgments. Importantly, each topic is typically represented by a single query. However, previous research has shown that queries created by users given a fixed information need may vary widely~\cite{bailey2017retrieval, zuccon2016query}. In the \uqv{}~\cite{bailey2016uqv100} dataset for instance, crowd workers on average created 57.7 unique queries for a given information need as instantiated as a backstory, e.g. \textit{``You have heard quite a lot about cheap computing as being the way of the future, including one recent model called a Raspberry Pi. You start thinking about buying one, and wonder how much they cost.''}

\definecolor{Gray}{gray}{0.9}
\begin{table}[!htb]
\small
\centering
\caption{Examples of BERT effectiveness drops (nDCG@10 $\Delta$) when we replace the original query from \trecdl{} by an automatic (except for the first two lines that were produced manually) query variation. We focus here on transformations that change the {\color{white}\colorbox{Peach}{query syntax}}, but not its {\color{white}\colorbox{gray}{semantics}}.}
\label{table:motivation}
\begin{tabular}{@{}p{2.5cm}p{3.3cm}r@{}}
\toprule
\textbf{Original Query} & \textbf{Query Variation} & \textbf{nDCG@10 $\Delta$} \\ \midrule
\rowcolor{Gray}
popular food in switzerland & popular food in zurich
{\color{white}\colorbox{gray}{\genspec{}}} &  \\ \midrule
\rowcolor{Gray}
cost of interior concrete flooring & concrete flooring finishing
{\color{white}\colorbox{gray}{\aspectchange}} &  \\ \midrule
what is theraderm used for & what is \textbf{thrraderm} used for 
{\color{white}\colorbox{Peach}{\misspelling}}& \textcolor{red}{-1.00} (-100\%) \\ \midrule
anthropological definition of environment & anthropological definition \textbf{\sout{of}} environment 
{\color{white}\colorbox{Peach}{\naturality{}}}
& \textcolor{red}{-0.15} (\phantom{0}-26\%) \\ 
\midrule
right pelvic pain causes & \textbf{causes} pelvic pain \textbf{right} 
{\color{white}\colorbox{Peach}{\ordering{}}}
& \textcolor{red}{-0.18} (\phantom{0}-46\%) \\ \midrule
define visceral & \textbf{what is} visceral 
{\color{white}\colorbox{Peach}{\paraphrase{}}}
& \textcolor{red}{-0.26} (\phantom{0}-38\%) \\ \bottomrule
\end{tabular}
\end{table}

We thus argue that it is necessary to investigate the robustness of retrieval pipelines in light of \emph{query variations} (i.e., different expressions of the same information need) that are \emph{likely to occur in practice}. That different query variations lead to vastly different ranking qualities is anecdotally shown in Table~\ref{table:motivation} for a vanilla BERT model for ranking~\cite{nogueira2019passage}. If, for example, the word order of the original query from \trecdl{} \emph{right pelvic pain causes} is changed to \emph{causes pelvic pain right}, the retrieval effectiveness of the resulting ranking drops by 46\%. Similarly, paraphrasing \emph{define visceral} to \emph{what is visceral} reduces the retrieval effectiveness by 38\%. 

In our work, we quantify the extent to which different retrieval models are susceptible to different types of query variations as measured by their drop in retrieval effectiveness. In contrast to prior works that either analyze behaviour of models when faced with modifications to the documents~\cite{macavaney2020abnirml}, analyze models through the lens of IR axioms~\cite{rennings2019axiomatic,camara2020diagnosing} or analyze NLP models via general natural language text adversarial examples~\cite{ribeiro2020beyond,gardner2020evaluating}, we instantiate our \emph{query variations} based on user-created data. Concretely, we manually label a large fraction of \uqv{} queries\footnote{To our knowledge, \uqv{} is the only publicly available dataset that contains a large number of query variations for a set of information needs.} and extract six types of frequently occurring query transitions: \genspec{}, \aspectchange{}, \misspelling{}, \naturality{}, \ordering{} and \paraphrase{}---an example of each is shown in Table~\ref{table:motivation}. The last four of these categories change the query syntax but not its semantics. For each of the syntax-changing categories, we develop automated approaches that enable us to generate query variations of each category for any input query. With these \emph{query variation generators} in place, we conduct extensive empirical work on the recent \trecdl{}~\cite{craswell2020overview} and \antique{}~\cite{hashemi2020antique} datasets to answer the following research question: \emph{Are retrieval pipelines robust to different variations in queries that do not change its semantics?} To this end we consider seven ranking approaches: two traditional lexical models (\bm{}~\cite{robertson1994some} and \bmrm{}~\cite{abdul2004umass}), two neural re-ranking approaches that do not make use of transformers (\KNRM{}~\cite{xiong2017end} and \CKNRM{}~\cite{dai2018convolutional}) and three transformer-based re-ranking approaches (\EPIC{}~\cite{macavaney2020expansion}, \BERT{}~\cite{nogueira2019passage} and \T{}~\cite{nogueira2020document}). Additionally, motivated by the fact that certain query variations can improve the retrieval effectiveness compared to using the original query~\cite{belkin1995combining,benham2019boosting}, we study the combination of automatic query variations with rank fusion~\cite{cormack2009reciprocal}. Our main findings are as follows:

\begin{itemize}[leftmargin=7pt]
    \item The four types of syntax-changing query variations differ in the extent to which they degrade retrieval effectiveness: \emph{misspellings} have the largest effect (with an average drop of 0.25 nDCG@10 points across seven retrieval models for \trecdl{}) while the \emph{word ordering} has the least effect (with an average drop of nDCG@10 smaller than 0.01 for \trecdl{}).
    \item Different types of ranking models make similar mistakes. For example, effectiveness decreases for models based on transformer language models are higher for \naturality{} query variations compared to decreases when using traditional lexical models.
    \item While rank fusion mitigates the drops in retrieval effectiveness when compared to using a single query variation, it does not achieve the full potential of the combination of query variations. An oracle that always select the best query achieves gains of 0.08 and 0.06 nDCG@10 points on \trecdl{} and \antique{} respectively.
\end{itemize}

Our work indicates that more research is required to improve the robustness of retrieval pipelines. Evaluation benchmarks should aim to have multiple query variations for the same information need in order to evaluate whether ranking pipelines are indeed robust, and we provide here a number of methods to automatically generate such query variations for any dataset.

\section{Related Work}
To put our work in context, we now describe prior research into query variations and then move on to research analyzing neural (IR) models.

\subsection{Query Variation}
A number of studies have argued that evaluation in IR tasks should take into account multiple instantiations of the same information need, i.e. query variations, due to their impact on the effectiveness of ranking models~\cite{spark1975report,belkin1993effect,buckley1999trec,bailey2015user,bailey2016uqv100,moffat2015pooled,zuccon2016query,bailey2017retrieval}. \citet{zuccon2016query} proposed a mean-variance framework to explicitly take into account query variations when comparing different IR systems. \citet{bailey2017retrieval} argued that a model should be consistent to different query variations, and proposed a measure of consistency which gives additional information to effectiveness measurements.

Besides a better evaluation of models, query variations can also be employed to improve the overall effectiveness of ranking models, for instance by combining the different rankings obtained from them~\cite{belkin1995combining,benham2019boosting} or by modelling relevance of multiple query variations~\cite{lu2019relevance}. They have also shown to been helpful for the problem of query performance prediction~\cite{zendel2019information}.

Different methods to automatically generate query variations have been proposed. \citet{benham2018towards} proposed to obtain query expansions through a relevance model which is built by issuing the original query against an external corpora and expanding it with additional terms from the set of external feedback documents. \citet{lu2019relevance} employed a query-url click graph and generated query variations automatically using a two-step backward walk process. \citet{chakraborty2020retrievability} generated query variations automatically based on an external knowledge base with a prior term distribution or by building a relevance model in a iterative manner. 

Our work differs from previous work on automatic query variation generation in the following ways: \one{} our methods do not require access to external corpora, a relevance model or a query-url click graph; \two{} we are not concerned with generating queries with the sole purpose of improving effectiveness, but in generating queries that are likely to occur in practice; and \three{} each of our generator methods follows a category of our taxonomy of query variations which allows us to \emph{diagnose} ranking models' effectiveness by analyzing what types of variations are more detrimental to what ranking models.

\subsection{Model Understanding}
The success of pre-trained transformer-based language models such as \BERT{}~\cite{devlin2019bert} and \T{}~\cite{raffel2019exploring} on several IR benchmarks---a comprehensive account of the effectiveness gains can be found in~\cite{lin2020pretrained}---has lead to research on understanding their behaviour and the reasons behind their significant gains in ranking effectiveness~\cite{qiao2019understanding, camara2020diagnosing, macavaney2020abnirml,zhan2020analysis,padigela2019investigating}. 

\citet{camara2020diagnosing} showed that \BERT{} does not adhere to IR axioms, i.e., heuristics that a reasonable IR model should fulfill, through the use of diagnostic datasets. \citet{macavaney2020abnirml} expanded on the axiomatic diagnostic datasets~\cite{rennings2019axiomatic} with ABNIRML, a framework to understand the behaviour of neural ranking models using three different strategies: measure and match (controlling certain measurements such as relevance or term frequency and changing another), manipulation of the documents' text (for example by shuffling words or replacing it with the query) and through the transfer of Natural Language Processing (NLP) datasets (for example comparing documents that are more/less fluent or formal with inferred queries). We expand on ~\citet{macavaney2020abnirml}'s work by proposing textual manipulations---unlike previous methods we are inspired by \emph{user-created} variations---to the queries instead of the documents and examine the robustness in terms of effectiveness of neural ranking models to such manipulations.

A different direction of research in NLP has challenged how well current evaluation schemes through the use of held-out test sets are actually evaluating the desired capabilities of the models. For example, \citet{gardner2020evaluating} proposed the manual creation of contrast sets---small perturbations that preserve artifacts but change the true label---in order to evaluate the models' decision boundaries for different NLP tasks. They showed that the model effectiveness on such contrast sets can be up to 25\% lower than on the original test sets. Inspired by behavioral testing, i.e. validating input output behaviour without knowledge about internal structure, from software engineering tests, \citet{ribeiro2020beyond} proposed to test NLP models with three different types of tests: minimum functionality tests (simple examples where the model should not fail), label (such as positive, negative and neutral in sentiment analysis) invariant changes to the input, and modifications to the input with known outcomes. With such tests at hand they were able to find actionable failures in different commercial NLP models that had already been extensively tested. It has also been shown that neural models developed for different NLP tasks can be tricked by adversarial examples~\cite{garg2020bae, alzantot2018generating, gao2018black}, i.e. examples with perturbations indiscernible by humans which get misclassified by the model. In terms of queries modifications, \cite{wu2021neural,zhuang2021dealing} found typos to be detrimental to the effectiveness of neural rankers. \citet{wu2021neural} analyzed the robustness of neural rankers with respect to three dimensions: difficult queries from similar distribution, out-of-domain cases, and defense against adversarial operations. Our work differs from the adversarial line of research by evaluating the robustness of models to query modifications that could be generated by humans, i.e. transformations that naturally occur, and not modifications optimized to trick neural models.



\begin{table*}[]
\small
\caption{Taxonomy of query variations derived from a sample of the \uqv{} dataset. Last column is the count of each query variation found on \uqv{} based on manual annotation of tuples of queries for the same information need. * spelling errors were already fixed for the \uqv{} pairs.}
\begin{tabular}{@{}lp{0.25\textwidth}p{0.09\textwidth}p{0.15\textwidth}lp{0.15\textwidth}l@{}}
\toprule
\textbf{Category} & \textbf{Definition} & \textbf{Changes Semantics} & 
\multicolumn{3}{l}{\textbf{\querypair{} Examples from \uqv{}}} & \textbf{Count (\%)} \\ \midrule
\rowcolor{Gray}
\Genspec{} & Generalizes or specializes within the same information need. & $\checkmark$  & american civil war & $\leftrightarrow$ & number of battles in south carolina during civil war & 172 (26.34\%) \\ \midrule
\rowcolor{Gray}
\Aspectchange{} & Moves between related but different aspects within the same information need. & $\checkmark$  & what types of spiders can bite you while gardening & $\leftrightarrow$ & signs of spider bite & 111 (17.00\%) \\ \midrule
\Misspelling{} & Adds or removes spelling errors. &  & raspberry pi & $\leftrightarrow$ & raspeberry pi &  * \\ \midrule
\Naturality{} & Moves between keyword queries and natural language queries. &  & how does zinc relate to wilson's disease & $\leftrightarrow$ & zinc wilson's disease & 118 (18.07\%)  \\ \midrule
\Ordering{} & Changes the order of words &  & carotid cavernous fistula treatment. & $\leftrightarrow$ & treatment carotid cavernous fistula & \phantom{0}37 (\phantom{0}5.67\%)\\ \midrule
\Paraphrase{} & Rephrases the query by modifying one or more words. &  & cures for a bald spot & $\leftrightarrow$ & cures for baldness & 215 (32.92\%)  \\ \bottomrule
\end{tabular}
\label{table:taxonomy}
\end{table*}
\section{Automatic Query Variations}

We now first describe in Section~\ref{sec:taxonomy} how we arrived at our query variation categories in a data-driven manner by annotating a large set of user-created query variations from \uqv{}.

We end up with six categories: four that change the syntax (but not the semantics) and two that change the semantics. \textbf{In our work, we focus on the four syntax-changing categories.} In Section~\ref{section:query_variation_methods} we subsequently describe our methods to automatically generate query variations for each category of the taxonomy that does not change the query semantics.

\subsection{UQV Taxonomy}\label{sec:taxonomy}
In order to better understand how queries differ when we compare different query variations for the same information need, we resort to analyzing variations from the \uqv{} dataset. \uqv{} contains query variations for 100 (sub)-topics from the TREC 2013 and 2014 web tracks, written by crowd workers who received a ``backstory'' for each topic as a starting point. On average, \uqv{} contains 57.7 spelling corrected (corrected by the \uqv{} authors using the spelling service of the Bing
search engine) query variations per topic. We consider a query variation pair \querypair{} to be a set of two queries $q_i$ and $q_j$ that were provided in \uqv{} for the same backstory. In total, 365K such pairs exist; Table~\ref{table:taxonomy} (4th column) contains a number of \querypair{} examples. We sampled 100 query variation pairs from the 365K available ones for manual annotation. Three authors of this paper (the ``annotators'') performed an open card sort~\cite{wood2008card}. The annotators independently sorted the query variation pairs into different piles and named them, each representing a transformation $T$ that can be applied to $q_i$ and then leads to $q_j$, i.e. $T(q_i) = q_j$. Multiple transformations might be applied to $q_i$ in order to yield $q_j$, e.g. $T_2(T_1(q_i)) = q_j$.

After the independent sorting step, the different piles were discussed and merged where necessary, which yielded five categories of transformations. Since the \uqv{} data used had already been spelling-corrected by its authors, we added the category \emph{misspellings}. The resulting taxonomy can be found in Table~\ref{table:taxonomy}. It contains a concrete definition and examples for each of our---in total---six categories: \one{} \emph{generalization or specialization}, \two{} \emph{aspect change}, \three{} \emph{misspelling}, \four{} \emph{naturality}, \five{} \emph{word ordering} and \six{} \emph{paraphrasing}. We observed two broad types of transformations: transformations that change the semantics of the query and transformations that do not change the semantics. The \genspec{} and \aspectchange{} transformations fall into the former type, whereas all other categories fall into the latter. We highlight here that unlike previous categorizations that describe how users revise queries in e-commerce~\cite{10.1007/978-3-030-72240-1_14,hirsch2020query}, how to generate better queries to substitute the original query~\cite{jones2006generating}, how users reformulate queries in a session~\cite{jansen2009patterns}, we study here how to categorize \emph{query variations} for the same information need which is a related but different problem.

Having arrived at our six categories, our annotators then labeled an additional set of 550 \querypair{} randomly sampled pairs from \uqv{} in order to determine the distribution of these categories in \uqv{}. Each \querypair{} was labelled as belonging to one (or more) of the five categories (with the exception of \misspelling{} which, as already stated, had already been corrected by the \uqv{} authors). In order to determine the inter-annotator agreement, 25 \querypair{} pairs were labelled by all three annotators, and 175 pairs were each labelled by a single annotator. The inter-annotator agreement~\cite{cohen1960coefficient} was moderate (Cohen's $\kappa=0.42$); the disagreements were highest for the \naturality{} and \paraphrase{} categories. We found that a total of 56 \querypair{} pairs had more than one category assigned to it\footnote{For example, the pair \{\textit{``what is doctor zhivago all about''}, \textit{``dr zhivago synopsis''}\} had both \paraphrase{} and \naturality{} labels, as it goes from a natural language question to a keyword-base question and also paraphrases \textit{``doctor [...] all about''} to \textit{``dr [...] synopsis}''}. The resulting distribution is shown in Table~\ref{table:taxonomy} (right-most column); the categories of query variations that change the query without changing its semantics account for ~57\% of all the transformations. In contrast, 43\% of query variations are semantic changes. Among the syntax-changing categories, we found \naturality{} to be the most common with 33\% of all transformations falling into this category.
Having observed that query variations change the syntax, but not the semantics for the majority of cases, \textbf{we focus in the remainder of our work on syntax-changing query variations}. We leave the exploration of query variation generators for \genspec{} and \aspectchange{} as future work.

\begin{table*}[ht!]
\small
\caption{Example of applying each method $M$ for the query `\textit{what is durable medical equipment consist of}' from \trecdl{} resulting in valid examples. Rightmost columns indicate the total percentage of valid queries by automatic query variation method based on manual annotation of queries from the test sets.}
\label{table:query_variation_methods}
\begin{tabular}{@{}llccc@{}}
\toprule
\textbf{Category} & \textbf{Method Name} & \multicolumn{1}{c}{\textbf{$M$(`\textit{what is durable medical equipment consist of}')}} & \multicolumn{1}{l}{\textbf{\trecdl{}}} & \multicolumn{1}{l}{\textbf{\antique{}}}  \\ \midrule
\multirow{3}{*}{\Misspelling{}} & \charswap{} & \textit{what is durable \textbf{mdeical} equipment consist of}   & 100.00\% & 99.50\% \\ \cmidrule(l){2-5} 
 & \charsub{} & \textit{what is durable \textbf{medycal} equipment consist of}   & 97.67\% & 91.00\% \\ \cmidrule(l){2-5} 
 & \charqwerty{} & \textit{what is durable medical equipment \textbf{xonsist} of}  & 97.67\% & 98.50\%  \\ \midrule
\multirow{2}{*}{\Naturality{}} & \removestop{} & \textit{\textbf{\sout{what is}} durable medical equipment consist \textbf{\sout{of}}}   & 86.05\% & 99.50\% \\ \cmidrule(l){2-5} 
 & \tfiverectotitle{} & \textit{\textbf{\sout{what is}} durable medical equipment}  \textbf{\sout{consist of}}  & 81.40\% & 68.00\% \\ \midrule
\Ordering{} & \orderswap{} & \textit{\textbf{medical} is durable \textbf{what} equipment consist of}   & 100.00\% & 100.00\%  \\ \midrule
\multirow{4}{*}{\Paraphrase{}} & \backtranslation{} & \textit{what is \textbf{sustainable} medical equipment} \sout{consist of}  & 53.49\% & 46.50\% \\ \cmidrule(l){2-5} 
 & \tfiveqqp{} & \textit{what is durable medical equipment \textbf{\sout{consist of}}}  & 60.47\% & 52.50\% \\ \cmidrule(l){2-5} 
 & \wordembedsyn{} & \textit{what is durable \textbf{medicinal} equipment consist of}   & 62.79\% & 62.00\% \\ \cmidrule(l){2-5} 
 & \wordnetsyn{} & \textit{what is \textbf{long lasting} medical equipment consist of}   & 37.21\% & 35.50\%  \\ \bottomrule
\end{tabular}
\end{table*}

\subsection{Query Generators} \label{section:query_variation_methods}

For each of the four syntax-changing categories, we explored different methods that generate query variations of the specified category. After an initial exploration of different query generator methods for each category, and filtering approaches that did not generate valid variations for the category and approaches that have high correlation with each other, we employed a total of ten different methods. These methods are listed in Table~\ref{table:query_variation_methods}, each with an example transformation. We explain each one in more detail in this section. A method $M_C$ receives as input a query $q$ and outputs a query variation $\hat{q}$: $M_C(q) = \hat{q}$. 

While most of the methods can generate multiple variations for a single input query (for example by replacing different words of the same query by synonyms or by including several spelling mistakes), for the experiments in the paper we resort to using a single query variation per method which already yields enough data for analysis (see $\S$ \ref{section:datasets}). Inspired by adversarial examples, we aim to make minimal perturbations to the input text when possible, e.g. replace only one word by a synonym, increasing the chances of obtaining valid variations.

\subsubsection{{\Misspelling{}}}
The three methods in this category add one spelling error to the query; the query term an error is introduced in is chosen uniformly at random.

\begin{description}[leftmargin=5pt]
    \item[\charswap{}] Swaps two neighbouring characters from a random query term (excluding stopwords\footnote{We use the NLTK english stopwords list for all the methods; it is available at \url{https://www.nltk.org/}.}).
    \item[\charsub{}] Replaces a random character from a random query term (excluding stopwords) with a randomly chosen new ASCII character.
    \item[\charqwerty{}] Replaces a random character of a random query term (excluding stopwords) with another character from the QWERTY keyboard such that only characters in close proximity are chosen, replicating errors that come from typing too quickly.
 \end{description} 
 
\subsubsection{{\Naturality{}}} The two methods in this category transform natural language queries into keyword queries.

\begin{description}[leftmargin=5pt]
    \item[\removestop{}] Removes all stopwords from the query.
    \item[\tfiverectotitle{}] Applies an encoder-decoder transformer model (here we employ T5~\cite{raffel2019exploring}) that we fine-tuned on the task of generating the title of a TREC topic title based on the TREC topic description. For example, a title and description tuple from \textit{`trec-robust04'}: `\textit{Evidence that rap music has a negative effect on young people.}' $\rightarrow$ `\textit{Rap and Crime}'. We collect pairs of title and description from eleven datasets available through the IR datasets library~\cite{macavaney:sigir2021-irds}: \texttt{trec-robust04}, \texttt{trec-tb-2004}, \texttt{aquaint/trec-robust-2005}, \texttt{gov/trec-web-2002}, \texttt{ntcir-www-2}, \texttt{ntcir-www-3},  \texttt{trec-misinfo-2019}, \texttt{cord19/trec-covid}, \texttt{dd-trec-2015}, \texttt{dd-trec-2016} and \texttt{dd-trec-2017}. Overall, we fine-tuned our model on 1322 description/title tuples. 
\end{description}

\subsubsection{{\Ordering{}}}
In this category, we employ only one basic method to shuffle words as done by previous research on the order of words~\cite{macavaney2020abnirml,pham2020out}.

\begin{description}[leftmargin=5pt]
    \item[\orderswap{}] Randomly swap two words of the query.
\end{description}

\subsubsection{{\Paraphrase{}}} 
The four methods in this category change one or more query terms in the process of paraphrasing. 

\begin{description}[leftmargin=5pt]
    \item[\backtranslation{}] Applies a translation method to the query to a pivot language, i.e. an auxiliary language, and from the pivot language back to the original language of the query, i.e. English. In our experiments we employ the M2M100~\cite{fan2020beyond} model, a multilingual model that can translate between any pair of 100 languages, and we use \textit{`German'} as the pivot language, which yielded better results---shown by manual inspection of the generated variations---than the other two languages for which the model had the most data for training (\textit{`Spanish'} and \textit{`French'}). This technique has been used before as a way to generate paraphrases~\cite{federmann2019multilingual,mallinson2017paraphrasing}.
    \item[\tfiveqqp{}] Applies an encoder-decoder transformer model (here we employ T5~\cite{raffel2019exploring}) that was fine-tuned on the task of generating a paraphrase question from the original question\footnote{As available here \url{https://huggingface.co/ramsrigouthamg/t5_paraphraser}}. The model employs the Quora Question Pairs\footnote{\url{https://www.kaggle.com/c/quora-question-pairs}} dataset for fine-tuning, which has 400k pairs of questions like the following: `\textit{How do you start a bakery?}' $\rightarrow$ `\textit{How can one start a bakery business?}'. We also tested T5 models fine-tuned for PAWS~\cite{pawsx2019emnlp} and the combination of PAWS and Quora Question Pairs, but the manual inspection of the generated queries revealed that T5 fine-tuned for Quora Question Pairs generated a higher number of valid variations.
    
    \item[\wordembedsyn{}] Replaces a non-stop word by a synonym as defined by the nearest neighbour word in the embedding space according to a counter fitted-Glove embedding which yields better synonyms than standard Glove embeddings~\cite{mrkvsic2016counter}. 
    \item[\wordnetsyn{}] Replaces a non-stop word by a the first synonym found on WordNet~\footnote{\url{https://wordnet.princeton.edu/}}. If there are no words with valid synonyms it will not output a valid variation.
\end{description}
        
\section{Experimental Setup}
In this section we describe our experimental setup aimed to answer the following research question: \emph{are retrieval pipelines robust to different variations in queries that do not change its semantics?}

\subsection{Datasets} \label{section:datasets}
We consider the following datasets in our experiments: \trecdl{}~\cite{craswell2020overview} for the passage retrieval task and \antique{}~\cite{hashemi2020antique} for non-factoid question answering task, containing 43 and 200 queries respectively in their test sets. For each of the test set queries, we generate one query variation for each of the proposed methods, and we use the manual annotation described in this section ($\S$\ref{section:qvannotation}) to take into account only the valid generated query variations in our experiments. The statistics of the datasets can be found in Table~\ref{table:dataset_stats}.

\begin{table}[]
\small
\caption{Statistics of the datasets.}
\label{table:dataset_stats}
\begin{tabular}{@{}lrr@{}}
\toprule
 & \textbf{\trecdl{}} & \textbf{\antique{} }\\ \midrule
\#Q train & 367013 & 2426 \\
\#Q valid & 5193 & - \\
\#Q test & 43 & 200 \\ \midrule
\# terms/ Q test & 5.51 & 10.51 \\
\# valid query variations & 334 & 1706 \\ \bottomrule
\end{tabular}
\end{table}

\subsection{Ranking Models}
We use different ranking models that cover from lexical traditional models (\trad{}) such as \bm{}, to neural ranking models (\nn{}) such as \KNRM{} and neural ranking models that employ transformer-based language models (\nnlm{}) such as \BERT{}. For all of our experiments, we apply \bm{} as a first stage retriever and re-rank the top 100 results with the neural ranking models, which is an established and efficient approach~\cite{lin2020pretrained}.

For \textbf{\bm{}}~\cite{robertson1994some} and \textbf{\bmrm{}}~\cite{abdul2004umass} we resort to the default hyperparameters and implementation provided by the PyTerrier toolkit~\cite{pyterrier2020ictir}. We trained the kernel-based ranking models \textbf{\KNRM{}}~\cite{xiong2017end} and \textbf{\CKNRM{}}~\cite{dai2018convolutional} on the training sets of \trecdl{} and \antique{} using default settings from the OpenNIR~\cite{macavaney:wsdm2020-onir} implementation. For the BERT-based methods \textbf{\EPIC{}}~\cite{macavaney2020expansion}, an efficiency focused model that encodes query and documents separately, and \textbf{\BERT{}}~\cite{nogueira2019passage}, also known as monoBERT, which concatenates query and the document and makes predictions based on the \texttt{[CLS]} token representation, we fine-tune the \texttt{bert-base-uncased} model for the train datasets. For \textbf{\T{}}~\cite{raffel2019exploring} we use the monoT5~\cite{nogueira2020document} implementation from PyTerrier T5 plugin \footnote{\url{https://github.com/terrierteam/pyterrier_t5}} which has the pre-trained weights for MSMarco~\cite{nguyen2016ms} by the original authors of monoT5.

\subsection{Query Generators Implementation}
As for our methods of generating query variations, for \tfiverectotitle{} and \tfiveqqp{} we rely on pre-trained T5 models (\texttt{t5-base}) and we fine-tune them using the Huggingface transformers library~\cite{wolf-etal-2020-transformers}. For \backtranslation{} we use the \texttt{facebook/m2m100\_418M} pre-trained model from the transformers library\footnote{\url{https://huggingface.co/facebook/m2m100_418M}}. For all other methods, we use the implementations from the TextAttack~\cite{morris2020textattack} library.

\subsection{Quality of Query Generators} \label{section:qvannotation}

Given the automatic nature of the methods we introduced, we need to evaluate their quality: how good are these methods at generating query variations that a user would also generate? 

To this end, we consider two properties of the generated queries: \one{} $\hat{q}$ maintains the same semantics as $q$, and \two{} the syntax difference between $q$ and $\hat{q}$ can be attributed to the category $C$. All pairs of $q$ and $\hat{q}= M(q)$ from the test sets of \trecdl{} (43 queries) and \antique{} (200 queries) for each of the 10 automatic variation methods went to the following process. First, we automatically set the variations from \misspelling{}\footnote{\misspelling{} methods can generate invalid queries when all words of the query are stop-words (e.g. `\textit{how is it being you}' from \antique{} would generate the same query as output since there is no non stop-words to modify)} and \ordering{} as valid, since they are rule based transformations to the input. Then all transformations that generate a variation that is identical to the input query ($\hat{q}= M(q) = q$) was automatically set to invalid. The annotators (the authors) then annotated independently the remaining 1371 pairs of $\{q, \hat{q}\}$ for the two mentioned properties (binary labels). The percentage of queries that are valid (both desired properties) are displayed at the right-most columns of Table~\ref{table:query_variation_methods} for the 10 automatic variation methods used in the paper and all combinations of $\{q, \hat{q}\}$ (2430). 

We find the methods in the \paraphrase{} category to yield the largest percentage of invalid query variations: fewer than 38\% of query variations generated via \wordnetsyn{} are valid. A manual inspection of the invalid queries reveal the following insights: \one{} \tfiverectotitle{} at times removes query terms that are important for the query and thus change its semantics (e.g. `\textit{if i had a bad breath what should i do}' $\rightarrow$ `\textit{if i had a}' ), \two{} \backtranslation{} and \tfiveqqp{} methods can generate an identical copy of the input query which was automatically labelled as invalid (e.g. `\textit{what is dark energy}' $\rightarrow$ `\textit{what is dark energy}') and \three{} transformations that replace words by their presumed synonyms (\wordembedsyn{} and \wordnetsyn{}) at times adds words that are not in fact synonymous in the query context (e.g. `\textit{what is dark energy}' $\rightarrow$ `\textit{what is blackness energy}' and `\textit{what is a active margin}' $\rightarrow$ `\textit{what is a active border}'). 

\textbf{To evaluate the robustness of the ranking models, we resort to using only the valid queries as defined by the manual annotations.} Overall, we have thus 2,040 valid queries for datasets \trecdl{} and \antique{} that we employ in the experiments that follow.    
\section{Results}
In this section we first describe our main results on the robustness of models to query variations, analyzing them by category of variation and by category of ranking model. We then move on to discussing the fusion of the ranking list obtained by the query variations.

\begin{table*}[]
\caption{Effectiveness (nDCG@10) of different methods for \trecdl{} and \antique{} when faced with different query variations. Bold indicates the highest values observed for each model and ${\downarrow}/{\uparrow}$ subscripts indicate statistically significant losses/improvements, using two-sided paired Student's T-Test at 95\% confidence interval when compared against the same model with the original queries. $\#Q$ indicates the number of valid query variations for the method (invalid query variations are replaced by the original query).}
\label{table:main_table_trec}
\begin{tabular}{@{}lllllllllr@{}}
\toprule
 &  & \multicolumn{7}{c}{\textbf{\trecdl{}}} & \multicolumn{1}{l}{} \\ \midrule
\textbf{Category} & \textbf{Query Variation} & \textbf{\bm{}} & \textbf{\bmrm{}} & \textbf{\KNRM{}} & \textbf{\CKNRM{}} & \textbf{\EPIC{}} & \textbf{\BERT{}} & \textbf{\T{}} & \textbf{\#Q} \\ \midrule
- & original query & \textbf{0.4795} & \textbf{0.5156} & \textbf{0.5015} & \textbf{0.4931} & \textbf{0.6240} & \textbf{0.6449} & 0.6997 & 43 \\ \hdashline
\multirow{3}{*}{\Misspelling{}} & \charswap{} & 0.2747$^{\downarrow}$ & 0.2748$^{\downarrow}$ & 0.3164$^{\downarrow}$ & 0.3085$^{\downarrow}$ & 0.3894$^{\downarrow}$ & 0.4161$^{\downarrow}$ & 0.4950$^{\downarrow}$ & 43 \\
 & \charsub{} & 0.2314$^{\downarrow}$ & 0.2333$^{\downarrow}$ & 0.2363$^{\downarrow}$ & 0.2263$^{\downarrow}$ & 0.2950$^{\downarrow}$ & 0.3283$^{\downarrow}$ & 0.3963$^{\downarrow}$ & 42 \\
 & \charqwerty{} & 0.2435$^{\downarrow}$ & 0.2504$^{\downarrow}$ & 0.2672$^{\downarrow}$ & 0.2965$^{\downarrow}$ & 0.3512$^{\downarrow}$ & 0.3867$^{\downarrow}$ & 0.4458$^{\downarrow}$ & 42 \\ \midrule
\multirow{2}{*}{\Naturality{}} & \removestop{} & 0.4778 & 0.5113 & 0.4842 & 0.4756 & 0.6214 & 0.6390 & 0.6866 & 37 \\
 & \tfiverectotitle{} & 0.4215 & 0.4344$^{\downarrow}$ & 0.3920$^{\downarrow}$ & 0.3928$^{\downarrow}$ & 0.5063$^{\downarrow}$ & 0.5361$^{\downarrow}$ & 0.5710$^{\downarrow}$ & 35 \\ \midrule
\Ordering{} & \orderswap{} & 0.4795 & 0.5156 & 0.5015 & 0.4708 & 0.6227 & 0.6349 & 0.6970 & 43 \\ \midrule
 \multirow{4}{*}{\Paraphrase{}} & \backtranslation{} & 0.3964$^{\downarrow}$ & 0.4195$^{\downarrow}$ & 0.3927$^{\downarrow}$ & 0.3605$^{\downarrow}$ & 0.5301$^{\downarrow}$ & 0.5467$^{\downarrow}$ & 0.6058$^{\downarrow}$ & 23 \\
 & \tfiveqqp{} & 0.4722 & 0.5043 & 0.4541$^{\downarrow}$ & 0.4609 & 0.6045 & 0.6396 & \textbf{0.7048} & 26 \\
 & \wordembedsyn{} & 0.3530$^{\downarrow}$ & 0.3539$^{\downarrow}$ & 0.3824$^{\downarrow}$ & 0.3680$^{\downarrow}$ & 0.4746$^{\downarrow}$ & 0.4721$^{\downarrow}$ & 0.5603$^{\downarrow}$ & 27 \\
 & \wordnetsyn{} & 0.3488$^{\downarrow}$ & 0.3650$^{\downarrow}$ & 0.3807$^{\downarrow}$ & 0.3605$^{\downarrow}$ & 0.4488$^{\downarrow}$ & 0.4474$^{\downarrow}$ & 0.5451$^{\downarrow}$ & 16 \\ \midrule
 &  & \multicolumn{7}{c}{\textbf{\antique{}}} & \multicolumn{1}{l}{} \\ \midrule
\textbf{Category} & \textbf{Query Variation} & \textbf{\bm{}} & \textbf{\bmrm{}} & \textbf{\KNRM{}} & \textbf{\CKNRM{}} & \textbf{\EPIC{}} & \textbf{\BERT{}} & \textbf{\T{}} & \textbf{\#Q} \\ \midrule
- & original query & \textbf{0.2286} & \textbf{0.2169} & 0.2175 & 0.2065 & 0.2663 & \textbf{0.4212} & \textbf{0.3338} & 200 \\  \hdashline
\multirow{3}{*}{\Misspelling{}} & \charswap{} & 0.1559$^{\downarrow}$ & 0.1478$^{\downarrow}$ & 0.1590$^{\downarrow}$ & 0.1451$^{\downarrow}$ & 0.1841$^{\downarrow}$ & 0.2868$^{\downarrow}$ & 0.2514$^{\downarrow}$ & 199 \\ 
 & \charsub{} & 0.1623$^{\downarrow}$ & 0.1593$^{\downarrow}$ & 0.1558$^{\downarrow}$ & 0.1476$^{\downarrow}$ & 0.1887$^{\downarrow}$ & 0.2797$^{\downarrow}$ & 0.2486$^{\downarrow}$ & 182 \\
 & \charqwerty{} & 0.1613$^{\downarrow}$ & 0.1527$^{\downarrow}$ & 0.1602$^{\downarrow}$ & 0.1550$^{\downarrow}$ & 0.1922$^{\downarrow}$ & 0.2987$^{\downarrow}$ & 0.2664$^{\downarrow}$ & 197 \\ \midrule
\multirow{2}{*}{\Naturality{}} & \removestop{} & 0.2270 & 0.2160 & \textbf{0.2222} & \textbf{0.2153} & \textbf{0.2693} & 0.3830$^{\downarrow}$ & 0.3200$^{\downarrow}$ & 199 \\
 & \tfiverectotitle{} & 0.1673$^{\downarrow}$ & 0.1646$^{\downarrow}$ & 0.1601$^{\downarrow}$ & 0.1669$^{\downarrow}$ & 0.2000$^{\downarrow}$ & 0.2695$^{\downarrow}$ & 0.2397$^{\downarrow}$ & 136 \\ \midrule
\Ordering{} & \orderswap{} & 0.2286 & 0.2168 & 0.2178 & 0.1978$^{\downarrow}$ & 0.2665 & 0.4134$^{\downarrow}$ & 0.3248$^{\downarrow}$ & 200 \\ \midrule
 \multirow{4}{*}{\Paraphrase{}}& \backtranslation{} & 0.1618$^{\downarrow}$ & 0.1546$^{\downarrow}$ & 0.1602$^{\downarrow}$ & 0.1438$^{\downarrow}$ & 0.2036$^{\downarrow}$ & 0.3045$^{\downarrow}$ & 0.2584$^{\downarrow}$ & 93 \\
 & \tfiveqqp{} & 0.2201 & 0.2065 & 0.2095 & 0.1962 & 0.2614 & 0.3926$^{\downarrow}$ & 0.3214$^{\downarrow}$ & 105 \\
 & \wordembedsyn{} & 0.1759$^{\downarrow}$ & 0.1719$^{\downarrow}$ & 0.1902$^{\downarrow}$ & 0.1690$^{\downarrow}$ & 0.2142$^{\downarrow}$ & 0.3245$^{\downarrow}$ & 0.2828$^{\downarrow}$ & 124 \\
 & \wordnetsyn{} & 0.1791$^{\downarrow}$ & 0.1751$^{\downarrow}$ & 0.1957$^{\downarrow}$ & 0.1765$^{\downarrow}$ & 0.2117$^{\downarrow}$ & 0.3244$^{\downarrow}$ & 0.2733$^{\downarrow}$ & 71 \\ \bottomrule
\end{tabular}
\end{table*}

\begin{figure}[]
    \centering
    \includegraphics[width=.45\textwidth]{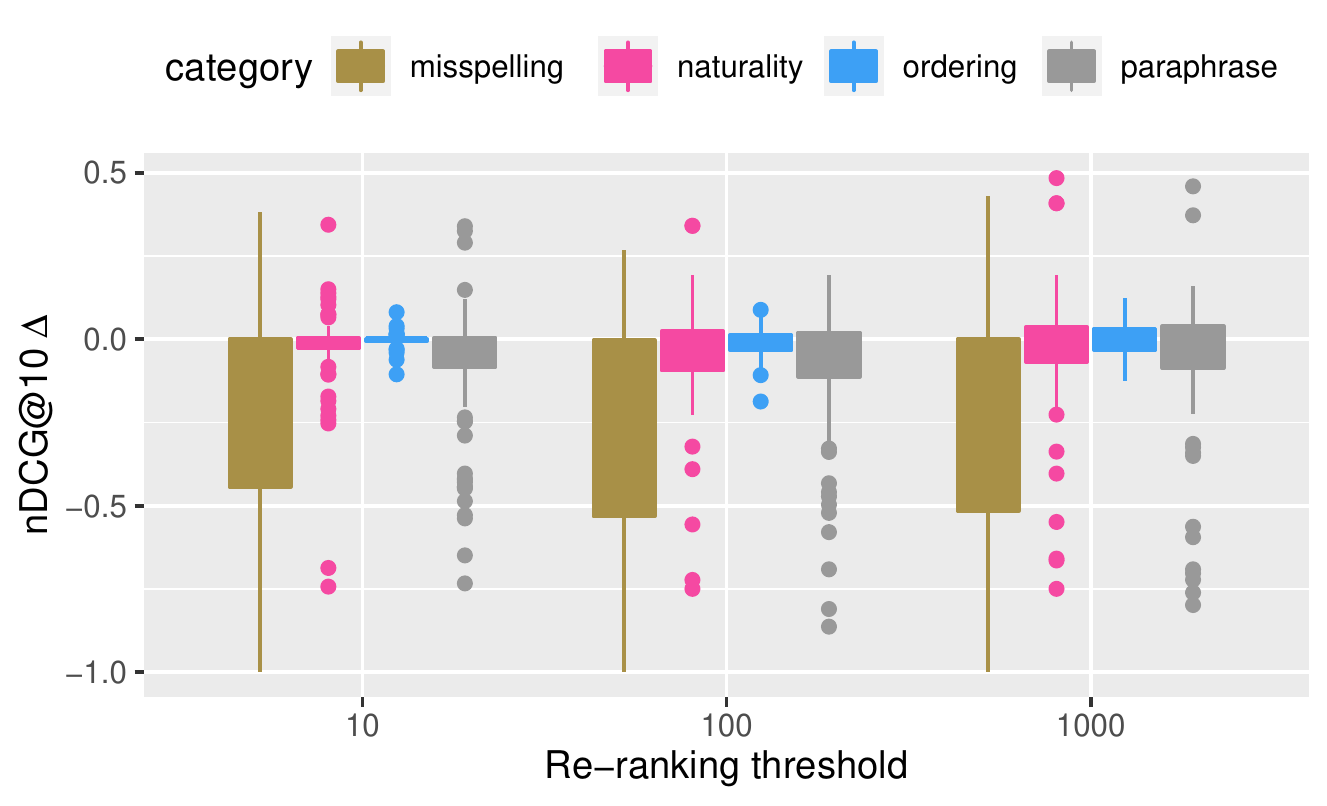}
    \caption{Distribution of nDCG@10 $\Delta$ for different re-ranking thresholds when using \BERT{} as a re-ranker.}
    \label{fig:reranking_threshold}
\end{figure}

\begin{figure*}[]
    \centering
    \includegraphics[width=.90\textwidth]{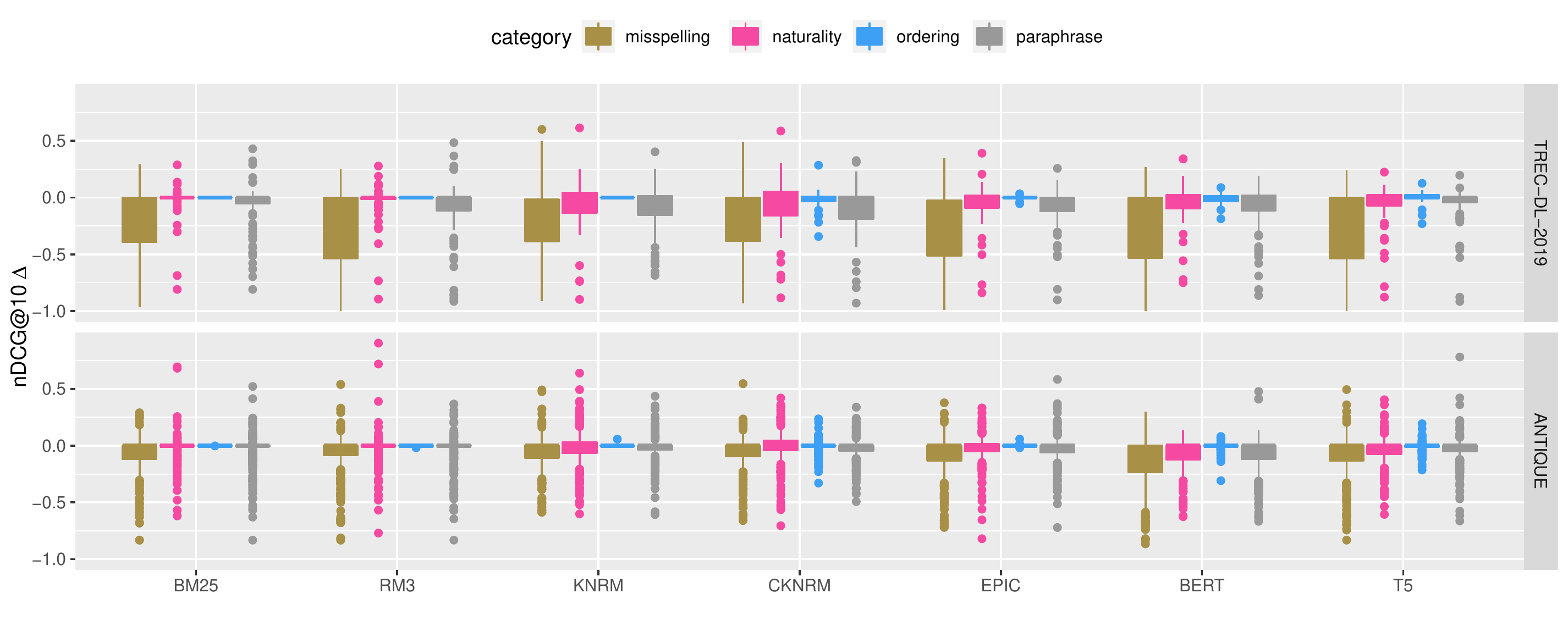}
    \caption{Distribution of nDCG@10 $\Delta$ when replacing the original query by the methods of each category.}
    \label{fig:delta_per_category}
\end{figure*}

\begin{figure}[]
    \centering
    \includegraphics[width=.45\textwidth]{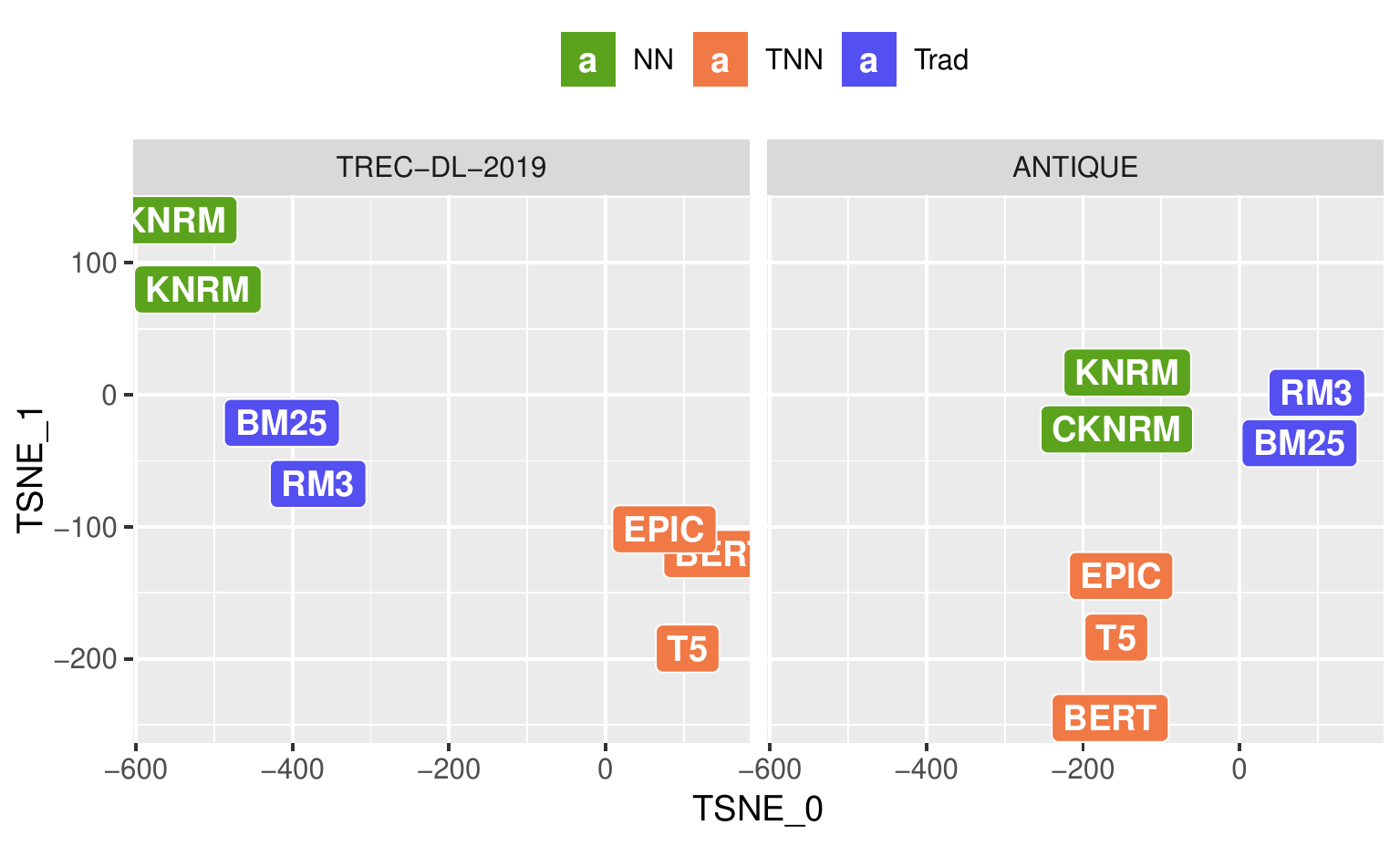}
    \caption{tSNE dimensionality reduction where each model is represented by the nDCG@10 $\Delta$ values obtained for each query and variation method ($\#Q \times \#M$).}
    \label{fig:tsne}
\end{figure}

\begin{table*}[]
\small
\caption{Pearson correlation between the nDCG@10 $\Delta$ of ranking models when we replace the original query by query variations of different categories for the \antique{} dataset.}
\label{table:correlation}
\begin{tabular}{@{}lllllrrrlllllrrr@{}}
\toprule
\multicolumn{8}{c}{\textbf{\Misspelling{}}} & \multicolumn{8}{c}{\textbf{\Naturality{}}} \\ \midrule
 & \textcolor{myblue}{\bm{}} & \textcolor{myblue}{\bmrm{}} & \textcolor{mygreen}{\KNRM{}} & \textcolor{mygreen}{\CKNRM{}} & \multicolumn{1}{l}{\textcolor{myorange}{\EPIC{}}} & \multicolumn{1}{l}{\textcolor{myorange}{\BERT{}}} & \multicolumn{1}{l}{\textcolor{myorange}{\T{}}} &  & \textcolor{myblue}{\bm{}} & \textcolor{myblue}{\bmrm{}} & \textcolor{mygreen}{\KNRM{}} & \textcolor{mygreen}{\CKNRM{}} & \multicolumn{1}{l}{\textcolor{myorange}{\EPIC{}}} & \multicolumn{1}{l}{\textcolor{myorange}{\BERT{}}} & \multicolumn{1}{l}{\textcolor{myorange}{\T{}}} \\ \midrule
\textcolor{myblue}{\bm{}} & \multicolumn{1}{r}{\cellcolor[HTML]{E67C73}1.00} & \multicolumn{1}{r}{\cellcolor[HTML]{EC9891}0.88} & \multicolumn{1}{r}{\cellcolor[HTML]{FBE7E6}0.54} & \multicolumn{1}{r}{\cellcolor[HTML]{FBE8E7}0.54} & \cellcolor[HTML]{F9DFDD}0.58 & \cellcolor[HTML]{FEF6F5}0.48 & \cellcolor[HTML]{FCEFEE}0.51 & \textcolor{myblue}{\bm{}} & \multicolumn{1}{r}{\cellcolor[HTML]{E67C73}1.00} & \multicolumn{1}{r}{\cellcolor[HTML]{EA8D86}0.90} & \multicolumn{1}{r}{\cellcolor[HTML]{FFFBFA}0.25} & \multicolumn{1}{r}{\cellcolor[HTML]{FEF9F8}0.26} & \cellcolor[HTML]{F9DDDB}0.43 & \cellcolor[HTML]{FCECEB}0.34 & \cellcolor[HTML]{FEFAF9}0.25 \\
\textcolor{myblue}{\bmrm{}} &  & \multicolumn{1}{r}{\cellcolor[HTML]{E67C73}1.00} & \multicolumn{1}{r}{\cellcolor[HTML]{FCEDEC}0.52} & \multicolumn{1}{r}{\cellcolor[HTML]{FBEAE9}0.53} & \cellcolor[HTML]{F9DFDD}0.58 & \cellcolor[HTML]{FFFFFF}0.44 & \cellcolor[HTML]{FEF7F7}0.47 & \textcolor{myblue}{\bmrm{}} &  & \multicolumn{1}{r}{\cellcolor[HTML]{E67C73}1.00} & \multicolumn{1}{r}{\cellcolor[HTML]{FFFFFF}0.22} & \multicolumn{1}{r}{\cellcolor[HTML]{FFFAFA}0.25} & \cellcolor[HTML]{FAE3E1}0.39 & \cellcolor[HTML]{FDF4F3}0.29 & \cellcolor[HTML]{FFFCFC}0.24 \\ \hdashline
\textcolor{mygreen}{\KNRM{}} &  &  & \multicolumn{1}{r}{\cellcolor[HTML]{E67C73}1.00} & \multicolumn{1}{r}{\cellcolor[HTML]{F2BAB5}0.74} & \cellcolor[HTML]{F5C9C5}0.67 & \cellcolor[HTML]{F8D9D6}0.60 & \cellcolor[HTML]{F5C6C2}0.68 & \textcolor{mygreen}{\KNRM{}} &  &  & \multicolumn{1}{r}{\cellcolor[HTML]{E67C73}1.00} & \multicolumn{1}{r}{\cellcolor[HTML]{F2BAB5}0.63} & \cellcolor[HTML]{F5CAC6}0.54 & \cellcolor[HTML]{FAE1DF}0.40 & \cellcolor[HTML]{FAE2E0}0.40 \\
\textcolor{mygreen}{\CKNRM{}} &  &  &  & \multicolumn{1}{r}{\cellcolor[HTML]{E67C73}1.00} & \cellcolor[HTML]{F7D2CF}0.63 & \cellcolor[HTML]{FBE8E7}0.54 & \cellcolor[HTML]{F9DDDB}0.59 & \textcolor{mygreen}{\CKNRM{}} &  &  &  & \multicolumn{1}{r}{\cellcolor[HTML]{E67C73}1.00} & \cellcolor[HTML]{F8D7D4}0.46 & \cellcolor[HTML]{FBE9E7}0.35 & \cellcolor[HTML]{FAE4E2}0.39 \\ \hdashline
\textcolor{myorange}{\EPIC{}} &  &  &  &  & \cellcolor[HTML]{E67C73}1.00 & \cellcolor[HTML]{F5CAC6}0.67 & \cellcolor[HTML]{F2BAB5}0.74 & \textcolor{myorange}{\EPIC{}} &  &  &  &  & \cellcolor[HTML]{E67C73}1.00 & \cellcolor[HTML]{F7D3D0}0.48 & \cellcolor[HTML]{F7D1CD}0.50 \\
\textcolor{myorange}{\BERT{}} &  &  &  &  & \multicolumn{1}{l}{} & \cellcolor[HTML]{E67C73}1.00 & \cellcolor[HTML]{F0ACA6}0.80 & \textcolor{myorange}{\BERT{}} &  &  &  &  & \multicolumn{1}{l}{} & \cellcolor[HTML]{E67C73}1.00 & \cellcolor[HTML]{F4C5C1}0.57 \\
\textcolor{myorange}{\T{}} &  &  &  &  & \multicolumn{1}{l}{} & \multicolumn{1}{l}{} & \cellcolor[HTML]{E67C73}1.00 & \textcolor{myorange}{\T{}} &  &  &  &  & \multicolumn{1}{l}{} & \multicolumn{1}{l}{} & \cellcolor[HTML]{E67C73}1.00 \\ \midrule
\multicolumn{8}{c}{\textbf{\Ordering{}}} & \multicolumn{8}{c}{\textbf{\Paraphrase{}}} \\ \midrule
 & \textcolor{myblue}{\bm{}} & \textcolor{myblue}{\bmrm{}} & \textcolor{mygreen}{\KNRM{}} & \textcolor{mygreen}{\CKNRM{}} & \multicolumn{1}{l}{\textcolor{myorange}{\EPIC{}}} & \multicolumn{1}{l}{\textcolor{myorange}{\BERT{}}} & \multicolumn{1}{l}{\textcolor{myorange}{\T{}}} &  & \textcolor{myblue}{\bm{}} & \textcolor{myblue}{\bmrm{}} & \textcolor{mygreen}{\KNRM{}} & \textcolor{mygreen}{\CKNRM{}} & \multicolumn{1}{l}{\textcolor{myorange}{\EPIC{}}} & \multicolumn{1}{l}{\textcolor{myorange}{\BERT{}}} & \multicolumn{1}{l}{\textcolor{myorange}{\T{}}} \\ \midrule
\textcolor{myblue}{\bm{}} & \multicolumn{1}{r}{\cellcolor[HTML]{E67C73}1.00} & \multicolumn{1}{r}{\cellcolor[HTML]{E67C73}1.00} & \multicolumn{1}{r}{\cellcolor[HTML]{FCEFEE}-0.01} & \multicolumn{1}{r}{\cellcolor[HTML]{FCEEED}0.01} & \cellcolor[HTML]{FCEEED}0.00 & \cellcolor[HTML]{FDF5F4}-0.06 & \cellcolor[HTML]{FBE8E7}0.06 & \textcolor{myblue}{\bm{}} & \multicolumn{1}{r}{\cellcolor[HTML]{E67C73}1.00} & \multicolumn{1}{r}{\cellcolor[HTML]{EB938C}0.88} & \multicolumn{1}{r}{\cellcolor[HTML]{FFFBFB}0.32} & \multicolumn{1}{r}{\cellcolor[HTML]{FFFBFB}0.32} & \cellcolor[HTML]{F9DDDB}0.48 & \cellcolor[HTML]{FBEAE9}0.41 & \cellcolor[HTML]{FEFAFA}0.32 \\
\textcolor{myblue}{\bmrm{}} &  & \multicolumn{1}{r}{\cellcolor[HTML]{E67C73}1.00} & \multicolumn{1}{r}{\cellcolor[HTML]{FCEFEE}-0.01} & \multicolumn{1}{r}{\cellcolor[HTML]{FCEEED}0.01} & \cellcolor[HTML]{FCEEED}0.00 & \cellcolor[HTML]{FDF5F4}-0.06 & \cellcolor[HTML]{FBE8E7}0.06 & \textcolor{myblue}{\bmrm{}} &  & \multicolumn{1}{r}{\cellcolor[HTML]{E67C73}1.00} & \multicolumn{1}{r}{\cellcolor[HTML]{FEF9F9}0.33} & \multicolumn{1}{r}{\cellcolor[HTML]{FFFFFF}0.29} & \cellcolor[HTML]{F8DAD7}0.50 & \cellcolor[HTML]{FBE8E7}0.42 & \cellcolor[HTML]{FEF5F4}0.35 \\ \hdashline
\textcolor{mygreen}{\KNRM{}} &  &  & \multicolumn{1}{r}{\cellcolor[HTML]{E67C73}1.00} & \multicolumn{1}{r}{\cellcolor[HTML]{FBE6E5}0.07} & \cellcolor[HTML]{FCEDEB}0.02 & \cellcolor[HTML]{FDF0EF}-0.01 & \cellcolor[HTML]{FDF2F1}-0.03 & \textcolor{mygreen}{\KNRM{}} &  &  & \multicolumn{1}{r}{\cellcolor[HTML]{E67C73}1.00} & \multicolumn{1}{r}{\cellcolor[HTML]{F5CAC6}0.58} & \cellcolor[HTML]{F8D6D3}0.52 & \cellcolor[HTML]{FFFBFB}0.32 & \cellcolor[HTML]{FBE6E4}0.43 \\
\textcolor{mygreen}{\CKNRM{}} &  &  &  & \multicolumn{1}{r}{\cellcolor[HTML]{E67C73}1.00} & \cellcolor[HTML]{FCEFEE}0.00 & \cellcolor[HTML]{FCEDEC}0.01 & \cellcolor[HTML]{FCEDEC}0.01 & \textcolor{mygreen}{\CKNRM{}} &  &  &  & \multicolumn{1}{r}{\cellcolor[HTML]{E67C73}1.00} & \cellcolor[HTML]{FBE9E7}0.42 & \cellcolor[HTML]{FEF5F4}0.35 & \cellcolor[HTML]{FBEAE9}0.41 \\ \hdashline
\textcolor{myorange}{\EPIC{}} &  &  &  &  & \cellcolor[HTML]{E67C73}1.00 & \cellcolor[HTML]{FFFFFF}-0.15 & \cellcolor[HTML]{FDF2F1}-0.03 & \textcolor{myorange}{\EPIC{}} &  &  &  &  & \cellcolor[HTML]{E67C73}1.00 & \cellcolor[HTML]{F7D4D1}0.53 & \cellcolor[HTML]{F4C5C1}0.61 \\
\textcolor{myorange}{\BERT{}} &  &  &  &  & \multicolumn{1}{l}{} & \cellcolor[HTML]{E67C73}1.00 & \cellcolor[HTML]{F9DCDA}0.16 & \textcolor{myorange}{\BERT{}} &  &  &  &  & \multicolumn{1}{l}{} & \cellcolor[HTML]{E67C73}1.00 & \cellcolor[HTML]{F3BBB7}0.66 \\
\textcolor{myorange}{\T{}} &  &  &  &  & \multicolumn{1}{l}{} & \multicolumn{1}{l}{} & \cellcolor[HTML]{E67C73}1.00 & \textcolor{myorange}{\T{}} &  &  &  &  & \multicolumn{1}{l}{} & \multicolumn{1}{l}{} & \cellcolor[HTML]{E67C73}1.00 \\ \bottomrule
\end{tabular}
\end{table*}

\subsection{Robustness to Query Variations}


In order to explore the robustness of our three types of ranking models (traditional, neural and transformer-based), we compare the effectiveness of our models when we replace the original query with the respective query variation. The results of this experiment are displayed in Table \ref{table:main_table_trec} for both the \trecdl{} and \antique{} datasets. Each row shows the effectiveness of the ranking models (columns) when using the queries obtained from each automatic query variation method. The last column ($\#Q$) displays the number of valid queries generated by each query variation method; the invalid queries are replaced with the original ones\footnote{While rows are directly comparable, methods with fewer valid queries are a lower bound of the potential decreases in effectiveness.}.

The results show that for most of the query variations and ranker combinations we observe a statistical significant effectiveness drop (49 out of 70 times for \trecdl{} and 54 out of 70 times for \antique{}), and that no set of query variations improves statistically over using the original query. If we look into the percentage of overall effectiveness decreases considering only the valid queries, we see on average that the models become 20.62\% and 19.21\% less effective for \trecdl{} and \antique{} respectively. \textbf{This answers our main research question indicating that retrieval pipelines are not robust to query variations}. This confirms previous empirical evidence that query variations induce a big variability effect on different IR systems~\cite{bailey2017retrieval,zuccon2016query}. We show that even with newer large-scale collections such as \trecdl{}, pipelines with neural ranking models are not robust to such variations.

There are several potential explanations for this drop in effectiveness besides the lack of robustness of neural rankers. The first-stage ranker may be the point of failure, being unable to retrieve sufficiently many relevant documents for the neural rankers to re-rank. It is also possible that the query variations lead to unjudged documents being ranked highly by the retrieval pipelines, which in the standard retrieval evaluation setup are considered non-relevant. We now present two experiments to show that these alternative explanations are not the cause in drop of retrieval effectiveness.

Let's focus first on the first-stage ranker. Figure~\ref{fig:reranking_threshold} shows the effect of increasing the re-ranking threshold on the distribution of nDCG@10 $\Delta$ when using \BERT{}, revealing that although the number of relevant documents on the re-ranking set increases (e.g. \bm{} has Recalls @10, @100 and @1000 on average of 0.06, 0.25 and 0.48 for \misspelling{} query variations), \BERT{} still struggles (negative $\Delta$) with query variations\footnote{Similar results are obtained for other neural rankers.}. This indicates that even if we increase the number of relevant documents in the list to be re-ranked, the re-rankers still fail when facing the query variations.

To further isolate the effect of the first-stage retrieval module, we analyzed whether the effectiveness of the pipelines would not degrade in case the first-stage retrieval was performed on the original query. In this experiment only the re-ranker models use the query variations and we check whether the effectiveness drops persist. The results reveal that there are still statistically significant effectiveness drops when only the re-ranker models use the query variations, although in smaller magnitude. While the drops in effectiveness of the pipelines when using query variations for the entire pipeline are on average of ~20\% in nDCG@10, when using the query variations only for re-ranking they are of ~9\%. \textbf{This indicates that not only the first stage retrieval module is not robust to query variations, but also the neural re-rankers}.

Let's now focus on the matter of unjudged documents. It is possible that we are underestimating the effectiveness of the retrieval pipelines when facing query variations if \one{} the number of unjudged documents in the top-10 ranked lists increases and \two{} they turn out to be relevant. When counting the amount of judged documents in the top-10 ranked lists of the retrieval pipelines, we find that on average the number actually increases (4.30\% for \trecdl{} and 0.36\% for \antique{}), \textbf{meaning that the performance drops of the retrieval pipelines can not be attributed to unjudged documents being brought up in the ranking by the query variations}.

\subsubsection{Robustness by Query Variation Category}

In order to study the effect of each query variation category, Figure~\ref{fig:delta_per_category} displays the nDCG@10 $\Delta$ (difference in effectiveness when replacing the original query by its variation) distribution per category and model. Although some queries variations have a positive effect (points with positive $\Delta$), the distributions are mostly skewed towards effectiveness decreases (negative $\Delta$). 

First we see that the on average the decreases are higher for the \misspelling{} category: -0.25 and -0.08 of nDCG@10 $\Delta$ for \trecdl{} and \antique{} respectively. We hypothesize that the effect is higher on \trecdl{} due to it having smaller queries than \trecdl{} (see average number of terms per query on Table~\ref{table:dataset_stats}).



The second highest effect on both datasets are the query variations from the \paraphrase{} category (-0.08 and -0.03 of nDCG@10 $\Delta$) followed by \naturality{} (-0.05 and -0.03). Compared to the \misspelling{} variations which in most cases degrade the effectiveness of our models, \paraphrase{} and \naturality{} have more queries for which the effect is positive, rendering the overall nDCG@10 $\Delta$ smaller.

Queries from the \ordering{} category have the least effect (less than 0.01). Since traditional methods are in fact bag-of-words models, changing the word order will not have any effect on them, which makes the average of all models' nDCG@10 $\Delta$ closer to zero. In the following section, we take a further look at how each type of ranking model is affected by each query variation method.

\subsubsection{Robustness by Model Category}

\begin{table*}[!htb]
\caption{Effectiveness (nDCG@10) of different methods when employing rank fusion (\rff{}) of the rankings obtained by using different sets of queries, e.g. $\rff{}_{\misspelling{}}$ fuses queries generated by \misspelling{} methods. Bold indicates the highest values observed for each model and ${\downarrow}/{\uparrow}$ subscripts indicate statistically significant losses/improvements, using t-test when compared against the same model with the original queries.}
\label{table:table_fusion}
\begin{tabular}{@{}llllllll@{}}
\toprule
 & \multicolumn{7}{c}{\textbf{\trecdl{}}} \\ \cmidrule(l){2-8} 
 & \textbf{\bm{}} & \textbf{\bmrm{}} & \textbf{\KNRM{}} & \textbf{\CKNRM{}} & \textbf{\EPIC{}} & \textbf{\BERT{}} & \textbf{\T{}} \\ \midrule
original query & 0.4795 & 0.5156 & 0.5015 & 0.4931 & 0.6240 & 0.6449 & 0.6997 \\ \hdashline
$\rff{}_{\Misspelling{}}$ & 0.3038$^{\downarrow}$ & 0.3092$^{\downarrow}$ & 0.3233$^{\downarrow}$ & 0.3175$^{\downarrow}$ & 0.3839$^{\downarrow}$ & 0.4164$^{\downarrow}$ & 0.4650$^{\downarrow}$ \\
$\rff{}_{\Naturality{}}$ & 0.4751 & 0.4972 & 0.4855 & 0.4639 & 0.5901 & 0.6161 & 0.6628 \\
$\rff{}_{\Paraphrase{}}$ & 0.4742 & 0.4865 & 0.4805 & 0.4330$^{\downarrow}$ & 0.5847 & 0.6120 & 0.6629 \\
$\rff{}_{All}$ & 0.4746 & 0.4976 & 0.5027 & 0.4951 & 0.5902$^{\downarrow}$ & 0.6031$^{\downarrow}$ & 0.6453$^{\downarrow}$ \\ \midrule
\oracle{} & \textbf{0.5401}$^{\uparrow}$ & \textbf{0.5774}$^{\uparrow}$ & \textbf{0.6055}$^{\uparrow}$ & \textbf{0.6121}$^{\uparrow}$ & \textbf{0.6990}$^{\uparrow}$ & \textbf{0.7194}$^{\uparrow}$ & \textbf{0.7595}$^{\uparrow}$ \\ \midrule
 & \multicolumn{7}{c}{\textbf{\antique{}}} \\ \cmidrule(l){2-8} 
 & \textbf{\bm{}} & \textbf{\bmrm{}} & \textbf{\KNRM{}} & \textbf{\CKNRM{}} & \textbf{\EPIC{}} & \textbf{\BERT{}} & \textbf{\T{}} \\ \midrule
original query & 0.2286 & 0.2169 & 0.2175 & 0.2065 & 0.2663 & 0.4212 & 0.3338 \\ \hdashline
$\rff{}_{\Misspelling{}}$ & 0.1712$^{\downarrow}$ & 0.1661$^{\downarrow}$ & 0.1758$^{\downarrow}$ & 0.1659$^{\downarrow}$ & 0.2060$^{\downarrow}$ & 0.2750$^{\downarrow}$ & 0.2435$^{\downarrow}$ \\
$\rff{}_{\Naturality{}}$ & 0.1847$^{\downarrow}$ & 0.1866$^{\downarrow}$ & 0.2029 & 0.2033 & 0.2406$^{\downarrow}$ & 0.3173$^{\downarrow}$ & 0.2707$^{\downarrow}$ \\
$\rff{}_{\Paraphrase{}}$ & 0.1909$^{\downarrow}$ & 0.1848$^{\downarrow}$ & 0.1917$^{\downarrow}$ & 0.1762$^{\downarrow}$ & 0.2380$^{\downarrow}$ & 0.3394$^{\downarrow}$ & 0.2882$^{\downarrow}$ \\
$\rff{}_{All}$ & 0.1998$^{\downarrow}$ & 0.1970$^{\downarrow}$ & 0.2150 & 0.2037 & 0.2430$^{\downarrow}$ & 0.3179$^{\downarrow}$ & 0.2729$^{\downarrow}$ \\ \midrule
\oracle{} & \textbf{0.2716}$^{\uparrow}$ & \textbf{0.2682}$^{\uparrow}$ & \textbf{0.2985}$^{\uparrow}$ & \textbf{0.2843}$^{\uparrow}$ & \textbf{0.3370}$^{\uparrow}$ & \textbf{0.4488}$^{\uparrow}$ & \textbf{0.3925}$^{\uparrow}$ \\ 
\bottomrule
\end{tabular}
\end{table*}

When we consider how different models are affected by the query variations, we see from Figure~\ref{fig:delta_per_category} that with the exception of \ordering{}, which has no effect on \bm{}, \bmrm{} and \KNRM{}, other transformations have a similar overall distribution of nDCG@10 $\Delta$ amongst different models. In order to understand if models (and category of models) make mistakes on the same queries, we label the models as follows: \bm{} and \bmrm{} are labelled as \textcolor{myblue}{\trad{}} (lexical matching), \KNRM{} and \CKNRM{} (neural network based) are labelled as \textcolor{mygreen}{\nn{}} and  \EPIC{}, \BERT{}, \T{} are labelled as \textcolor{myorange}{\nnlm{}} (transformer language model based). We then represent each model with the nDCG@10 $\Delta$ values obtained for each query and variation method resulting in a total of $\#Q \times \#M$ features per model. In order to visualize them we reduce this representation to 2 factors with tSNE ~\cite{van2008visualizing}, as shown in Figure~\ref{fig:tsne}. We observe that even though models have similar magnitudes and direction of nDCG@10 $\Delta$s, classes of models as indicated by color are clustered indicating that the query variations have similar effects for each type of model. 
We make a similar observation when we consider the correlation of nDCG@10 $\Delta$ amongst models for the different types of transformations as displayed in Table~\ref{table:correlation}, where there are groups with higher similarity that roughly correspond to the different types of models.

While \trad{} models have decreases of -0.03 (\trecdl{}) and -0.01 (\antique{}) nDCG@10 for \naturality{} query variations, the effect is higher on \nnlm{}: -0.05 and -0.04 respectively. This is evidence that neural ranking models based on heavily pre-trained language models have a slight preference for natural language queries as opposed to keyword queries, which is a finding aligned with previous work~\cite{dai2019deeper}. Another interesting finding is that the word order does not have a great effect on \nnlm{} models (decreases smaller than 0.01). This is in line with recent research that indicates that the word order might not be as important as initially thought for transformer models~\cite{pham2020out,sinha2021masked}.


\begin{figure}[]
    \centering
    \includegraphics[width=.45\textwidth]{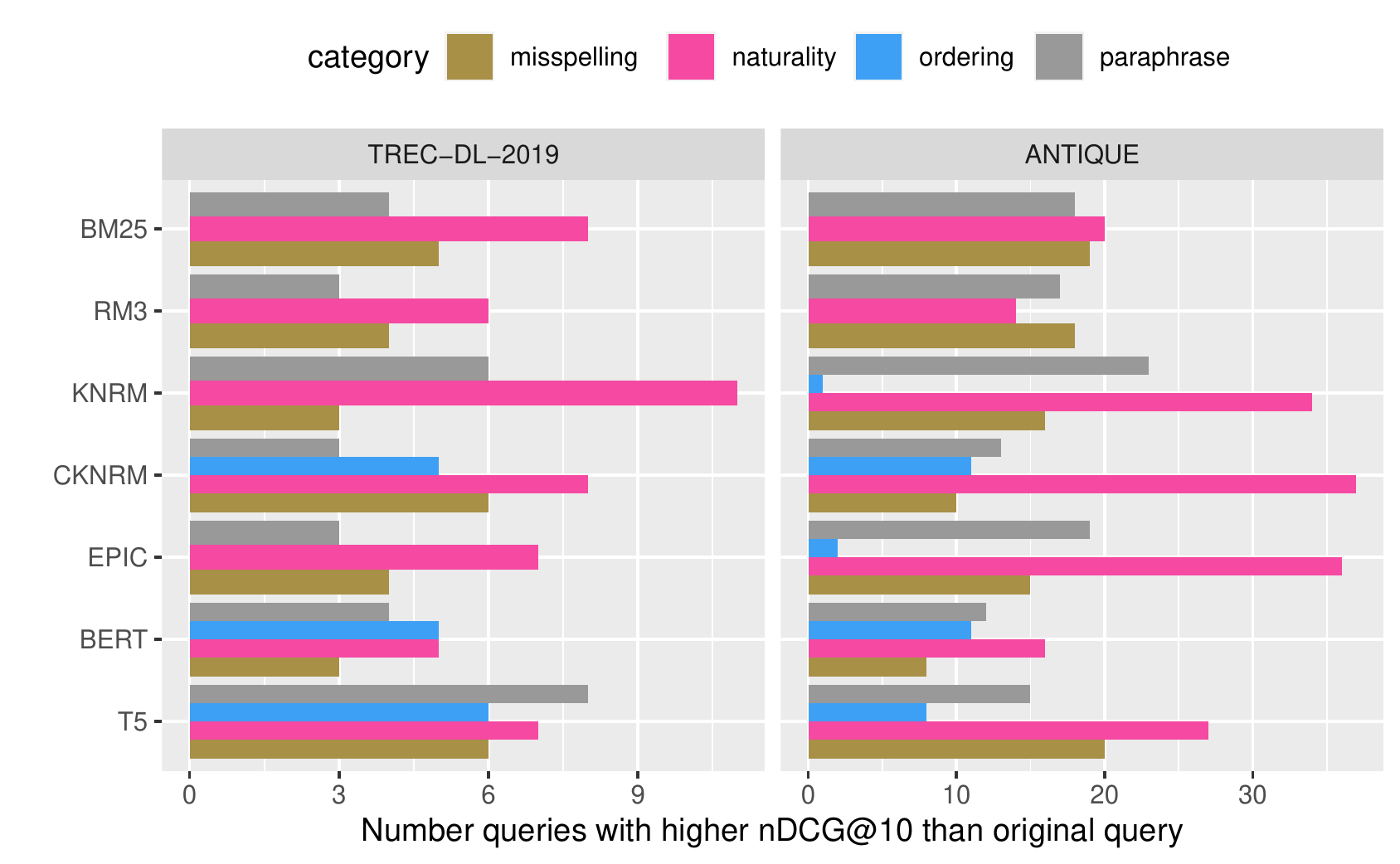}
    \caption{Distribution of query variations that are better than the original query by model and category of query variation.}
    \label{fig:best_query_dist}
\end{figure}

\subsection{Fusing Query Variations}

Although on average query variations make models less effective, there are cases when there are effectiveness gains (as shown with the positive nDCG@10 $\Delta$ in Figure~\ref{fig:delta_per_category}). This motivates the combination of different query variations to obtain better ranking effectiveness. In order to understand whether we can improve effectiveness of models by combining different query variations, we compare different methods for combining queries, as displayed in Table~\ref{table:table_fusion}. $\rff{}_{C}$ indicates that we fuse the results obtained from the query variations obtained after applying $M_C$ methods using the Reciprocal Rank Fusion (\rff{}) method~\cite{cormack2009reciprocal}, and $\rff{}_{All}$ fuses the results obtained by all query variation methods\footnote{\ordering{} was not included in the experiments as a separated row since it only has one method, but it is included in the $\rff{}_{All}$ method.}.

First we see that there is potential to have significant effectiveness gains, as shown by the last line (\oracle{}) where we always use the query with the highest retrieval effectiveness amongst query variations and the original query. The results show that combining query variations with \rff{} is better than using query variations individually (Table~\ref{table:main_table_trec}), and sometimes it is even the same as using the original query (no statistical difference). \textbf{Our results indicate that while rank fusion mitigates the decreases in effectiveness of different query variations ($\rff{}_{All}$ decreases are of 3\% and 10\% nDCG@10 for \trecdl{} and \antique{} respectively when compared to the original query), it does not improve the effectiveness over using the original query}. 

\subsubsection{When are query variations better?}
To better understand when models benefit from different query variations, we plot the distribution of query variations that improve over the original query by ranking model and query variation category in Figure~\ref{fig:best_query_dist}. We see that overall the queries obtained through the \naturality{} and \paraphrase{} methods are the ones that improve over the original queries the most. Intuitively, \paraphrase{} query variations can potentially rewrite the query with better terms (e.g. `\textit{why do criminals practice crime}' $\rightarrow$ `\textit{why do criminals practice misdemeanour}' \textcolor{mygreen}{+0.13} nDCG@10 for \BERT{} using \wordembedsyn{}), make queries grammatically correct (e.g. `\textit{how sun rises}' $\rightarrow$ `\textit{how does the sun rise}' \textcolor{mygreen}{+0.03} nDCG@10 for \BERT{} using \tfiveqqp{}) and also corrects spelling mistakes (e.g. `\textit{what is sosiology}' $\rightarrow$ `\textit{what is sociology}' \textcolor{mygreen}{+0.47} nDCG@10 for \BERT{} using \backtranslation{}). \naturality{} methods make the queries shorter (e.g. `\textit{who is robert gray}' $\rightarrow$ `\textit{robert gray}' \textcolor{mygreen}{+0.34} nDCG@10 for \BERT{} using \removestop{}), removing unnecessary information from the original query on certain cases.
    
\section{Conclusion}

In this work we studied the robustness of ranking models when faced with query variations. We first described a taxonomy of transformations between two queries for the same information need that characterizes how exactly a query is modified to arrive at one of its variants. We found six different types of transformations, and we focused our experiments on the ones that do not change the query semantics: \misspelling{}, \naturality{}, \ordering{} and \paraphrase{}. They account for 57\% of observed variations in the \uqv{} dataset. For each of these four categories we proposed different methods to automatically generate a query variation based on an input query. We studied the quality of the generated query variations, and based only on the valid ones we analyzed how robust retrieval pipelines are to them. Our experimental results on two different datasets quantify how much each model is affected by each type of query variation, demonstrating large effectiveness drops of 20\% on average when compared to the original queries from the test sets. We found rank fusion techniques to somewhat mitigate the drops in effectiveness. 

Our work highlights the need of creating test collections that include query variations to better understand model effectiveness. As future work, we believe that it is important to study \one{} how to automatically generate high quality query variation generators for categories that do change the semantics---while maintaining the same information need---of the query and \two{} techniques to improve the robustness of existing ranking pipelines.



\bibliographystyle{ACM-Reference-Format}
\bibliography{paper}

\end{document}